\documentclass[aip,jcp,reprint,groupedaddress]{revtex4-1}
\usepackage{graphicx}

\begin{document}

%\preprint{}

\title{Effects of strongly selective additives 
on volume phase transition in gels}


\author{Yuki Uematsu and Takeaki Araki}
%\thanks{}
\affiliation{Department of Physics, Kyoto University, Kyoto 606-8502, Japan}


\def\be{\begin{equation}}
\def\en{\end{equation}}
\def\bea{\begin{eqnarray}}
\def\ena{\end{eqnarray}}

\def\p{\partial}
\def\ep{\epsilon}
\def\gs{\gtrsim}
\def\ls{\lesssim} 
\def\ve{\varepsilon}
\def\n{\nabla}
\def\d{\delta}

\newcommand{\av}[1]{\langle{#1}\rangle}
\newcommand{\AV}[1]{\bigg \langle{#1}\bigg \rangle}
\newcommand{\bi}[1]{\mbox{\boldmath$#1$}}
\newcommand{\pp}[2]{\frac{\partial {#1}}{\partial {#2}}}
\newcommand{\ppp}[3]{{\bigg(}\frac{\partial {#1}}{\partial {#2}}{\bigg )}_{#3}}
\newcommand{\pppd}[3]{{\bigg(}\frac{d {#1}}{d {#2}}{\bigg )}_{#3}}
\newcommand{\pppm}[3]{{\bigg(}\frac{\delta{#1}}{\delta{#2}}{\bigg )}_{#3}}
\newcommand{\ten}[1]{\stackrel{\leftrightarrow}{\bi{#1}}}
\newcommand{\dis}[1]{[{#1}]_-^+}
\newcommand{\di}[1]{\nabla\cdot{{#1}}}
\newcommand{\digra}[2]{\nabla\cdot{#1}\nabla{#2}}

\date{\today}

\begin{abstract}
We investigate volume phase 
transition in gels immersed in mixture solvents, 
on the basis of a three-component Flory-Rehner theory. 
When the selectivity of the minority solvent component 
to the polymer network is strong, 
the gel tends to shrink with an increasing concentration of 
the additive, 
regardless of whether it is good or poor. 
This behavior 
originates from the difference of the 
additive concentration between inside and outside the gel. 
We also found the gap of the gel volume at the 
transition point can be controlled by adding the strongly selective solutes. 
By dissolving a strongly poor additive, for instance, 
the discontinuous volume phase transition can be extinguished. 
Furthermore, we observed that another volume phase trasition occurs 
far from the original transition point. 
These behaviors can be well explained by a simplified theory 
neglecting the non-linearity of the additive concentration. 

\end{abstract}

%\keywords{}

\maketitle

\section{introduction}

Swelling behavior of a polymer network was first investigated 
by Flory and Rehner \cite{Flory_JCP_1943}. 
On the basis of this work, discontinuous 
volume phase transition (VPT) of gels 
was predicted theoretically \cite{Dusek_JPSA_1968}. 
Thence, the VPT has received 
much attention from both scientific and industrial viewpoints 
\cite{Tanaka_PRL_1978,Tanaka_PRL_1980,Tanaka_SA_1981}. 
In various types of gels, 
addition of solutes into the solvent often plays very important roles 
in their volume change 
\cite{Tanaka_PRL_1978,Tanaka_PRL_1980,Tanaka_SA_1981,
Hirokawa_JCP_1984}. 
For example, the VPT was first realized 
in a polyacrylamide (PAA) gel immersed in a mixture 
of water and acetone \cite{Tanaka_PRL_1978}. 
In this experiment, 
the solvent quality, or $\chi$ parameter, is controlled by changing 
the volume fraction of the acetone. 
It is also well known that addition of salts dramatically 
affects the behaviors of ionic gels 
\cite{Ohmine_JCP_1982,Ricka_Macro_1984,FernandezNieves_JCP_2001}. 
Furthermore, a nonioninc hydrated gel such 
as poly($N$-isopropylacrylamide) gel 
(NIPA) changes its volume discontinuously. 
In the hydrated gels, the VPT is attributed to 
temperature-dependence of hydrogen bondings of the polymer network 
\cite{Hirokawa_JCP_1984,Amiya_JCP_1987}. 
It has been reported that its VPT is affected 
by adding salts \cite{Hirotsu_JCP_1987,Park_Macro_1993,
Inomata_Lang_1992,Annaka_JCP_2000} 
and other additives 
\cite{Hirotsu_JPSJ_1987,
Kokufuta_Macro_1993,
Kawasaki_JPC_1996,Sasaki_Macro_1997,Dhara_Lang_1999,
Koga_JPCB_2001,FTanaka_PRL_2008}. 

It is often assumed that the composition of a mixture solvent is 
constant in and out of a gel. 
This assumption is referred to as single liquid approximation (SLA). 
The SLA is experimentally confirmed in a NIPA gel 
in a dimethyl sulfoxide-water mixture, 
for instance \cite{Ishidao_CPS_1994}. 
However, the compositions of mixture solvents are not 
necessarily homogeneous 
\cite{Vasilevskaya_PSUSSR_1989,Iwatsubo_Macro_1995,
Ishidao_CPS_1994}. 
There, the difference of the composition between the 
interior and exterior 
of the gel would lead to a dramatic effect on its volume change. 
Some extended Flory-Huggins models dealing with 
a polymer network and two spieces of solvent molecules 
were developed to investigate the volume change 
of a polymer network in a mixture solvent 
\cite{Krigbaum_JPS_1954,Bristow_TFS_1959,
Vasilevskaya_PSUSSR_1989,Iwatsubo_Macro_1995,
Okeowo_Macro_2006}. 
In these theories, the volume change is characterized by the 
compositions of the three components and three $\chi$ parameters. 
They were studied systematically for fixed sets of the $\chi$ 
parameters. 
Usually, the $\chi$ parameters in polymer systems 
depend on environment parameters such as temperature. 
Since it is quite complicated to incorporate the temperature dependences 
of the $\chi$ parameters, 
the swelling behavior in mixture solvents has not been fully understood.

Recently, the effects 
of solutes with strong selective solvation on 
phase behaviors of a water-oil mixture was investigated 
\cite{Okamoto_PRE_2010,Onuki_COCIS_2011,Onuki_BCSJ_2011}. 
By dissolving a very small amount of strongly 
hydrophobic solute into the water-oil mixture, 
the oil-rich phase containing the solute is precipitated 
even in the one-phase region. 
This finding implies that such solutes would affect 
the VPT in gels.

The aim of this study is to clarify the effect of the additives 
on the VPT in gels, based on a simple theoretical 
argument. 
In particular, we focus on the cases, in which one of the solvent components 
has strong selectivity to the polymer network.

\section{model free energy}

Swelling behaviors of gels are well described by the Flory-Rehner theory 
\cite{Flory_JCP_1943}. 
The free energy $F$ consists of a mixing part $F_\mathrm{mix}$ 
and an elastic part $F_\mathrm{el}$ as
\begin{equation}
F=F_\mathrm{mix}+F_\mathrm{el}.
\end{equation}
For a gel immersed in a mixture solvent, 
the mixing free energy $F_\mathrm{mix}$ is given by 
\cite{
Krigbaum_JPS_1954,
Vasilevskaya_PSUSSR_1989,Iwatsubo_Macro_1995,
Okeowo_Macro_2006,Onuki_COCIS_2011} 

\begin{equation}
F_\mathrm{mix}=k_\mathrm{B} 
Tv_0^{-1}\left[V_\mathrm{g}f(\phi_\mathrm{1g},\phi_\mathrm{2g},\phi_3)
+V_\mathrm{s}f(\phi_\mathrm{1s},\phi_\mathrm{2s},0)\right], 
\label{eq:Fmix} 
\end{equation}
where $k_\mathrm{B}$ is the Boltzmann constant and $T$ is the temperature. 
$V_\mathrm{g}$ and $V_\mathrm{s}$ are the volumes of the 
interior and exterior of the gel, respectively. 
In this model, we assume that the volumes of each monomer 
of the polymer 
network and the solvent molecules are equal to 
a characteristic volume $v_0$ for simplicity. 
$f(\phi_1,\phi_2,\phi_3)$ is the 
Flory-Huggins type mixing free energy of 
the ternary system as \cite{Scott_JCP_1949}  
\begin{equation}
f(\phi_1,\phi_2,\phi_3)=\phi_1\ln\phi_1+\phi_2\ln\phi_2+\sum_{i<j}\chi_{ij}\phi_i\phi_j. 
\label{eq:FH} 
\end{equation}
Here, $\phi_1$ and $\phi_2$ are the volume fractions of the first and second 
components of the mixture solvent, respectively. 
In Eq.~(\ref{eq:Fmix}), $\phi_{i\mathrm{g}}$ and $\phi_{i\mathrm{s}}$ 
stand for the volume fraction of the $i$-th component ($i=1,2$) 
in and out of the gel. 
$\phi_3$ is the volume fraction of the polymer network. 
Since all the polymer chains are chemically connected 
forming a single network, 
the polymers are not dissolved outside the gel, that is, 
$\phi_{3{\mathrm{s}}}=0$. 
In Eq.~(\ref{eq:FH}), the first two terms 
stem from the translational entropy of the solvent molecules. 
Here, the translational entropy of the network is neglected. 
The last term is the interaction energy among the solvent molecules 
and the monomers of the network. 
$\chi_{ij}$ is the interaction parameter between the 
$i$-th and $j$-th components.

We employ Flory's rubber elasticity as \cite{Flory_book},
\begin{equation}
F_{\rm el}=\frac{1}{2}k_\mathrm{B}T\nu V_\mathrm{g0}
\left\{\sum_{
k=x,y,z}
\gamma
^2_k-3-2B\ln(\gamma_x\gamma_y\gamma_z)\right\}, 
\label{eq:Fel0}
\end{equation}
where $V_\mathrm{g0}$ is the volume of the gel in a reference state, 
$\nu$ is the density of cross linking points, 
and $\gamma_k$ is the elongation ratio in the $k$-axis ($k=x,y,z$). 
$B$ is a nonlinear elastic coefficient, 
which is often assumed $B=1/2$. 
In polyelectrolyte gels, the translational entropy of the 
counter-ions is renormalized into $B$ as $B+b$, 
where $b$ is the number 
of the dissociable monomers per effective chain. 
In this study, however, we do not consider the 
electric charges explicitly. 
In the case of isotropic swelling, 
the elongation ratio is coupled with the volume fraction 
of the polymer network as 
\begin{eqnarray}
\gamma_k=\left(\frac{V_{\rm g}}{V_{\rm g0}}\right)^{1/3}
=\left(\frac{\phi_{30}}{\phi_3}\right)^{1/3}, 
\end{eqnarray}
where $\phi_{30}$ is the volume fraction of the network 
in the reference state. 
Therefore, the elastic energy Eq.~(\ref{eq:Fel0}) 
is rewritten by a function of $\phi_3$ as, 
\begin{equation}
F_\mathrm{el}=\frac{1}{2}k_\mathrm{B}T\nu V_\mathrm{g0}
\left\{3\left(\frac{\phi_{30}}{\phi_3}\right)^{2/3}
-2B\ln\left(\frac{\phi_{30}}{\phi_3}\right)\right\}.
\end{equation}
We impose the incompressible conditions 
for both in and out of the gel, 
\begin{eqnarray}
&\phi_\mathrm{1g}+\phi_\mathrm{2g}+\phi_3=1, &\\
&\phi_\mathrm{1s}+\phi_\mathrm{2s}=1.&
\end{eqnarray}

Here, we define a grand potential as 
\begin{equation}
\Omega=F_\mathrm{mix}+F_\mathrm{el}
-\mu_2(V_\mathrm{g}\phi_\mathrm{2g}+V_\mathrm{s}\phi_\mathrm{2s})
+\kappa (V_{\rm g}+V_{\rm s}), 
\end{equation}
where $\mu_2$ and $\kappa$ are 
Lagrange multipliers to 
conserve the amount of the second component and the 
total volume. 
The equilibrium state is characterized by minimizing $\Omega$, 
\bea
&&\frac{\partial \Omega}{\partial \phi_{\rm 2g}}=
\frac{\partial \Omega}{\partial \phi_{\rm 2s}}=0.  
\label{eq:G1}\\
&&\frac{\partial \Omega}{\partial V_{\rm g}}
=\frac{\partial \Omega}{\partial V_{\rm s}}=0,
\label{eq:G2}
\ena
Then, we obtained the equilibrium conditions:  
\bea
&&\tilde{\mu}(\phi_{\rm 2g},\phi_3)=
\tilde{\mu}(\phi_{\rm 2s},0)
\left(=\frac{v_0\mu_2}{k_{\rm B}T}\right), 
\label{eq:mu}\\
&&\frac{k_{\rm B}T}{v_0}
\left(\tilde{\Pi}_0+\tilde{\Pi}_{\rm add}+\tilde{\Pi}_{\rm el}\right)=0, 
\label{eq:Pi}
\ena
where 
$\tilde{\mu}$ is the reduced chemical potential given by 
\bea
\tilde{\mu}(\phi_2,\phi_3)
&=& \ln\frac{\phi_2}{1-\phi_2-\phi_3}
\nonumber\\
&+&\chi_{12}(1-2\phi_2-\phi_3)+(\chi_{23}-\chi_{31})\phi_3. 
\label{eq:mu2}
\ena
Eq.~(\ref{eq:mu}) and (\ref{eq:Pi}) represent the 
balances of the chemical potential for the second component 
and the osmotic pressure, respectively. 

$\tilde{\Pi}_0$ is a part of the osmotic pressure for the gel 
without the second solvent component. 
It stems from the mixing free energy and given by 
\bea
\tilde{\Pi}_0=-\phi_3-\chi_{31}\phi_3^2-\ln(1-\phi_3). 
\label{eq:Pi0}
\ena
$\tilde{\Pi}_{\rm el}$ is the contribution of the elasticity as 
\bea
\tilde{\Pi}_{\rm el}=-\nu v_0\left\{
\left(\frac{\phi_3}{\phi_{30}}\right)^{1/3}
-B\left(\frac{\phi_3}{\phi_{30}}\right)\right\}. 
\label{eq:Piel}
\ena
If the solvent does not contain the additive $\phi_2=0$, 
the balance of the osmotic pressure is expressed by 
$\tilde{\Pi}_0+\tilde{\Pi}_{\rm el}=0$. 

$\tilde{\Pi}_{\rm ad}$ is the contribution of the 
second component and is given by 
\bea
\tilde{\Pi}_{\rm ad}
&&=-\ln\left(1-\frac{\phi_{\rm 2g}}{1-\phi_3}\right)
+\ln (1-\phi_{\rm 2s})
\nonumber\\
&&-\chi_{12}(\phi_{\rm 2g}^2
-\phi_{\rm 2s}^2)+G\phi_{\rm 2g}\phi_3, 
\label{eq:Piad0}
\ena
where $G=\chi_{23}-\chi_{31}-\chi_{12}$ is a parameter 
describing the affinity of the second component to the 
polymer network. 
$\tilde{\Pi}_0$ and $\tilde{\Pi}_{\rm add}$ come from the 
mixing free energey $F_{\rm mix}$. 
If $G<0$ and $|G|\gg 1$, the additive tends to be adsorbed selectively 
to the polymer network. 
If $G\gg 1$, on the other hand, 
the additive would be expelled from the gel. 

In this study, we assume that the binary solvent is 
completely mixed outside the gel.  
We note that 
the volume of the equilibirated gel does not change 
if we add more mixture solvent whose composition is equal to the 
equilibrated outer solvent. 
Hence, we take a limit of $V_{\rm s}/V_{\rm g}\rightarrow \infty$ 
with fixing $\phi_{\rm 2s}$, 
so that $\phi_{\rm 2s}$ is uniquely determined for a fixed $\mu_2$. 
Hereafter, we use $\phi_{\rm 2s}$ as a control parameter.

\section{Results and Discussions} 

\subsection{The first volume phase transition} 

We numerically study the effects of additives on volume changes of 
gels. 
First, we focus on the cases, 
in which the gel can undergo VPT without addtives. 
We set $\phi_{30}=5.0\times 10^{-2}$, 
$\nu v_0=1.0\times 10^{-2}$ and $B=0.75$ \cite{Tanaka_PRL_1978}. 

We vary $\chi_{31}$ continuously to induce the volume change 
with fixing the other parameters. 
We obtain the swellng curves of the gel 
by solving Eqs.~(\ref{eq:mu}) and (\ref{eq:Pi}) numerically. 
Figure 1 shows the swelling curves 
for several values of $\phi_{\rm 2s}$. 
In the absence of additives, 
the VPT occurs at 
$\chi_{\rm 31}=\chi_{\rm 31t}^{(0)}(\approx 0.93)$.
In Fig.~1 (a), 
we dissolve the additive of $\chi_{12}=0.0$ 
and 
$\chi_{23}=-1.0$
, which has affinity to 
the polymer network (pro-gel). 
As its concentration $\phi_{\rm 2s}$ is increased, 
the transition point shifts to higher $\chi_{31}$. 
In Fig.~1 (b), 
the additive of $\chi_{12}=0.0$ and 
$\chi_{23}=2.0$ is 
dissolved into the solvent. 
This additive dislikes the polymer network (anti-gel). 
It is shown that 
the transition point shifts to lower $\chi_{31}$ with 
an increasing $\phi_{\rm 2s}$. 
For the cases as in Fig.~1, 
the SLA can explain 
the effects of the solute on the volume changes qualitatively. 
In the SLA, 
the volume fractions of the solvent components inside 
the gel are assumed to be $\phi_{\rm 1g}=(1-\phi_3)(1-\phi_{\rm 2s})$ 
and $\phi_{\rm 2g}=(1-\phi_3)\phi_{\rm 2s}$. 
Then, the interaction parameter between the polymer network and 
the mixture solvent is approximated as 
\bea
\tilde{\chi}_{31}&=&\chi_{31}(1-\phi_{\rm 2s})+\chi_{23}\phi_{\rm 2s}
-\chi_{12}\phi_{\rm 2s}(1-\phi_{\rm 2s})\nonumber\\
&=&\chi_{31}+G\phi_{\rm 2s}+\mathcal{O}(\phi_{\rm 2s}^2). 
\label{eq:chi31}
\ena
This means that the interaction parameter changes effectively with 
$\phi_{\rm 2s}$ depending on $G$. 
For the plotted range in Fig.~1 (a) 
($\chi_{31}\in [0.92,0.95]$), 
$G$ remains negative, 
so that the solvent becomes more good, swelling the gel with an 
increasing $\phi_{\rm 2s}$.  
As shown in Fig.~1 (b), on the other hand, 
the solvent changes to more poor and 
the gel shrinks with $\phi_{\rm 2s}$ for the solute of positive $G$.

\begin{figure}
\includegraphics[width=84.5mm]{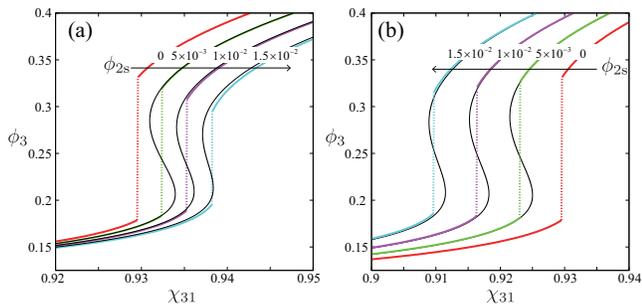}
\caption{
The swelling curves of a gel immersed in a mixture solvent. 
We set $\phi_{30}=5.0\times 10^{-2}$, $\nu v_0=1.0\times 10^{-2}$, 
$B=0.75$, and $\chi_{12}=0.0$. 
In the absence of additives, 
the gel undergoes volume phase transition with increasing $\chi_{31}$. 
We dissolve solutes of $\chi_{23}=-1.0$ (pro-gel) in (a) and 
$\chi_{23}=2.0$ (anti-gel) in (b). 
The black solid lines are calculated from Eq.~(\ref{eq:chi31_2}). 
}

\label{fig1}
\end{figure}

In Fig.~1, the absolute value of the resulting $G$ 
is rather small ($|G|\lesssim 2$). 
Next, we study the effects of solutes of strong selectivity $|G|\gg 1$. 
Figures~2 (a) and (b) show 
the swelling curves in the mixture solvents of 
$\chi_{23}=-9.0$ and $\chi_{23}=11.0$
, respectively. 
The other parameters are the same as those in Fig.~1. 
In the both cases, 
the transition points are shifted to lower $\chi_{31}$, 
regardless of whether the solute is good or poor. 
This behavior is in contrast to the volume changes in Fig.~1 
and indicates the SLA does not work well when the additives have the 
strong selectivity to the polymer network. 

\begin{figure}
\includegraphics[width=84.5mm]{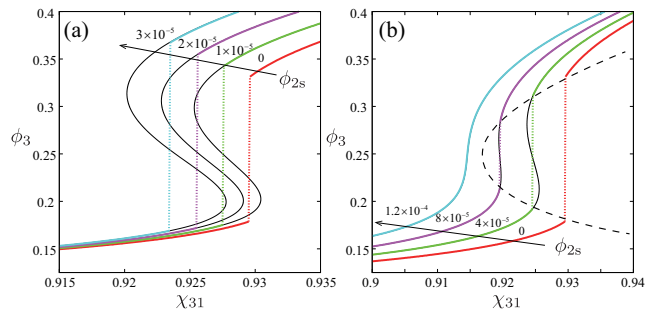}
\caption{
The swelling curves of a gel immersed in a mixture solvent. 
We set $\phi_{30}=5.0\times 10^{-2}$, $\nu v_0=1.0\times 10^{-2}$, 
$B=0.75$, and $\chi_{12}=0.0$. 
We dissolve solutes of $\chi_{23}=-9.0$ (pro-gel) in (a) 
and 
$\chi_{23}=11.0$ (anti-gel) in (b).
The black solid lines are calculated from Eq.~(\ref{eq:chi31_2}).
The additives have strong selectivities to the polymer network. 
}
\label{fig2}
\end{figure}

Figure~3 (a) shows the dependences of 
the transition shift  
$\Delta \chi_{\rm 31t}$ on the additive 
concentration $\phi_{\rm 2s}$, 
where $\Delta \chi_{\rm 31t}=\chi_{\rm 31t}
-\chi_{\rm 31t}^{(0)}$ and 
$\chi_{\rm 31t}^{(0)}$ is the transition point 
without additives. 
An increase of $\Delta \chi_{\rm 31t}$ 
is observed for 
$\chi_{23}\approx -1.0$, 
whereas $\Delta \chi_{\rm 31t}$ is 
lowered with an increasing $\phi_{\rm 2s}$ for the 
other additives. 
In the plotted range of $\phi_{\rm 2s}$, 
$\Delta \chi_{\rm 31t}$ has linear dependence on 
$\phi_{\rm 2s}$. 
The transition shifts for $\phi_{\rm 2s}=0.01$ 
are plotted with respect to $\chi_{23}$ in Fig.~3 (b). 
It indicates a non-monotonic behavior of the transition point. 

\begin{figure}
\includegraphics[width=85mm]{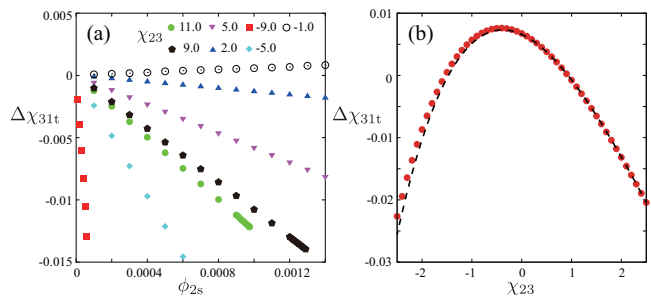}
\caption{
(a) The shifts of the trantision point from $\chi_{\rm 31t}^{(0)}$ 
are shown for several values of $\chi_{23}$. 
The shift is numerically obtained with the same parameters 
as in Figs.~1 and 2. 
For $\chi_{23}=9.0$ and $11.0$, 
the transition points are terminated 
since the discontinous volume changes disappear. 
(b) The dependence of $\Delta \chi_{\rm 31t}$ on $\chi_{23}$ 
is plotted for $\phi_{\rm 2s}=0.01$. 
The broken curve represents $g(\phi_{\rm 3c},\bar{G})$ [Eq. (\ref{eq:g})] 
with $\phi_\mathrm{3c}=0.239$ and 
$\chi_\mathrm{31t}^{(0)}=0.93$. 
}
\label{fig3}
\end{figure}

The difference of the additive effects between 
$\chi_{23}=-1.0$ and $\chi_{23}=-9.0$ is notable, 
since the both addivies have tendencies to be adsorbed 
onto the polymer network. 
When $|\chi_{23}|$ is small, the difference 
between $\phi_{\rm 2g}$ and $\phi_{\rm 2s}$ is 
rather small. 
As explained by Eq.~(\ref{eq:chi31}), 
the addition of the solute with negative $\chi_{23}$ 
makes the mixture solvent more good to the polymer. 
Thus, the gel is simply swollen with an increasing $\phi_{\rm 2s}$. 
If $|\chi_{23}|$ is large enough, on the other hand, 
the mixture solvent inside the gel becomes much denser 
than that outside the gel. 
By shrinking its volume, the polymer network tends to 
increase the contact points to the additive molecules. 
This non-linear effect would give rise to 
the difference between 
$\chi_{23}=-1.0$ and $\chi_{23}=-9.0$.

Figure~2 also shows that 
the gap of the volume transition 
depends on the additive concentration 
when the additive has strong selectivity. 
As $\phi_{\rm 2s}$ is increased, the volume gap is increased 
for $\chi_{23}=-9.0$ [see Fig.~2 (a)], 
while it is decreased for $\chi_{23}=11.0$ [Fig.~2 (b)]. 
For $\chi_{23}=11.0$, 
in particular, the gap disappears 
eventually at a certain concentration 
$\phi_{\rm 2s}=\phi_{\rm 2st}$,
above which the gel does not undergo the VPT [see below]. 
Figure~4 plots the gap of the volume fraction 
$\Delta \phi_{\rm 3t}(=\phi_{\rm 3t+}-\phi_{\rm 3t-})$  
at the transition point, 
where $\phi_{\rm 3t+}$ and $\phi_{\rm 3t-}$ are the 
volume fractions just below and above the transition point. 
For a large positive value of $\chi_{23}$, 
$\Delta \phi_{\rm 3t}$ decreases toward zero with 
an increasing $\phi_{\rm 2s}$. 

\begin{figure}
\includegraphics[width=42.5mm]{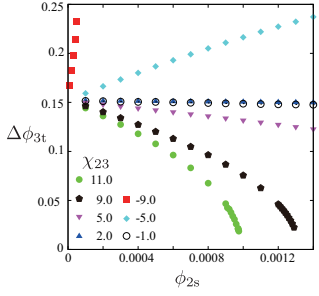}
\caption{
The gap of the volume fraction $\phi_3$ at 
the transition is plotted with $\phi_{\rm 2s}$. 
We set $\phi_{30}=5.0\times 10^{-2}$, $\nu v_0=1.0\times 10^{-2}$, 
$B=0.75$, and $\chi_{12}=0.0$ as in Figs.~1 and 2. 
}
\label{fig4}
\end{figure}

It is known that the elasticity parameter $B$ 
influences the nature of the VPT \cite{Tanaka_PRL_1978}. 
Figure~5 shows the swelling curves of a gel of 
$B=0.72$ 
with varying $\chi_{31}$. 
In the absence of additive, 
the volume of this gel changes continuously. 
By dissolving the pro-gel additive of $\chi_{23}=-9.0$, 
the swelling curve becomes non-monotonic with 
respect to $\chi_{31}$, 
so that the gel undergoes the VPT. 
Figures~2 (b) and 5 indicate that 
we can arbitrarily induce or erase the VPT by adding 
solutes with the strong selectivities. 

\begin{figure}
\includegraphics[width=42.5mm]{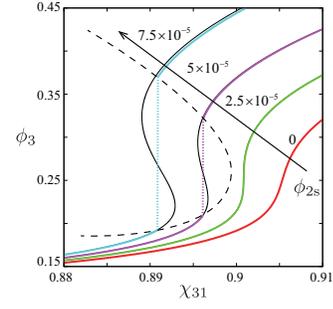}
\caption{
The swelling curves of a gel immersed in a mixture solvent. 
We set $\phi_{30}=5.0\times 10^{-2}$, $\nu v_0=1.0\times 10^{-2}$, 
$B=0.72$, and $\chi_{12}=0.0$. 
We dissolve the pro-gel solute of $\chi_{23}=-9.0$. 
The black solid lines are calculated from Eq.~(\ref{eq:chi31_2}) .
When $\phi_{\rm 2s}=0$, the gel changes its volume 
continuously without VPT. 
}
\label{fig5}
\end{figure}

We assume that only a subtle amount of the additive is 
dissolved in the outer solvent, 
{\it i.e.}, $\phi_{\rm 2s} \ll 1$. 
From Eq.~(\ref{eq:mu}), the volume fraction of the additive inside the 
gel is expressed by 
\bea
\phi_{\rm 2g}=(1-\phi_3)e^{-G\phi_3}\phi_{\rm 2s}
+\mathcal{O}(\phi_{\rm 2s}^2)
\label{eq:phi2g}. 
\ena
Substituting Eq.~(\ref{eq:phi2g}) into Eq.~(\ref{eq:Piad0}), 
we obtain 
\bea
\tilde{\Pi}_{\rm ad}&=& \ln (1-\phi_{\rm 2s})
-\ln(1-\phi_{\rm 2s} e^{-G\phi_3})
\nonumber\\
&&+G\phi_3(1-\phi_3)e^{-G\phi_3}\phi_{\rm 2s}
+\mathcal{O}(\phi_{\rm 2g}^2)\nonumber\\
&=& \left[e^{-G\phi_3}\{1+G\phi_3(1-\phi_3)\}-1\right]
\phi_{\rm 2s}+\mathcal{O}(\phi_{\rm 2g}^2). \nonumber\\
\label{eq:Piad}
\ena
Then, the swelling curve 
is approximatedly given from Eqs.~(\ref{eq:Pi0}), (\ref{eq:Piel}) 
and (\ref{eq:Piad}) by 
\bea
\chi_{31}\cong -\phi_3^{-2}\ln (1-\phi_3)-\alpha \phi_3^{-5/3}+(\beta-1)
\phi_3^{-1} \nonumber \\
+[e^{-G\phi_3}\{1+G\phi_3(1-\phi_3)\}-1]\phi_3^{-2}\phi_{\rm 2s}, 
\label{eq:chi31_2}
\ena
where $\alpha =\nu v_0/\phi_{30}^{1/3}$ and 
$\beta =\nu v_0 B/ \phi_{30}$. 
Since the right hand side of Eq.~(\ref{eq:chi31_2}) includes 
$\chi_{31}$ via $G$, we cannot obtain an analytical solution of 
$\chi_{31}$. 
Then, we replace $G$ to 
$\bar{G}=\chi_{23}-\chi_{\rm 31t}^{(0)}-\chi_{21}$ 
and regard $\bar{G}$ as a fixed parameter. 
The approximated curves of Eq.~(\ref{eq:chi31_2}) 
are drawn in Figs.~1, 2 and 5. 
They are well in agreement 
with the numerical solutions.

Regarding $\beta$ as a variable, 
a tri-critical point of the VPT is given by 
\bea
\left.\frac{\partial\chi_{31}}{\partial\phi_3}
\right|_{\phi_{\rm 3c},\beta_{\rm c}}=0, \quad
\left. 
\frac{\partial^2\chi_{31}}{\partial\phi_3^2}
\right|_{\phi_{\rm 3c},\beta_{\rm c}}=0 
\label{eq:tri}
\ena
Here, $\alpha$ is treated to be fixed. 
By solving Eq.~(\ref{eq:tri}), 
we obtain $\phi_{\rm 3c}$ and $\beta_{\rm c}$, 
which give the swelling curve passing through the tri-critical 
point. 
In the vicinity of the tri-critical point, 
the swelling curve is approximated by 
\bea
\chi_{31}\approx \chi_{\rm 31t}-
\frac{\beta-\beta_{\rm c}}{\phi_{\rm 3c}^2}(\phi_3-\phi_{\rm 3c})
+\frac{u}{6}(\phi_3-\phi_{\rm 3c})^3, 
\label{eq:critical} 
\ena
where $u$ is a positive constant. 
If $\beta<\beta_{\rm c}$, 
the curve of $\chi_{31}$ changes monotonically with $\phi_3$, 
such that the gel volume changes continuously with changing $\chi_{31}$. 
On the other hand, inflection points appears when $\beta>\beta_{\rm c}$. 
If so, the gel becomes mechanically unstable, 
so that it exhibits the discontinuous volume changes 
as $\Delta \phi_{\rm 3t}=2\sqrt{(\beta-\beta_{\rm c})/u}/\phi_{\rm 3c}$ 
at $\chi_{31}=\chi_{\rm 31t}$
\cite{Tanaka_PRL_1978}.

The part of the osmotic pressure $\tilde{\Pi}_{\rm ad}$ 
shifts the tri-critical point in 
Eq.~(\ref{eq:critical}) 
as 
\bea
\chi_{\rm 31t}&=&\chi_{\rm 31t}^{(0)}+g(\phi_{\rm 3c},\bar{G})\phi_{\rm 2s},\\
\beta_{\rm c}
&=&\beta_{\rm c}^{(0)}+h(\phi_{\rm 3c},\bar{G})\phi_{\rm 3c}^2\phi_{\rm 2s}. 
\ena
Here, $\beta_{\rm c}^{(0)}$ gives the tri-critical 
point in the case of $\phi_{\rm 2s}=0$. 
From Eq.~(\ref{eq:chi31_2}), 
the prefactors $g$ and $h$ are given by 
\bea
g(\phi_3,G)&=&\phi_{\rm 3}^{-2}[
e^{-G\phi_{\rm 3}}\{1+G\phi_{\rm 3}(1-\phi_{\rm 3})\}-1],
\label{eq:g}\\
h(\phi_3,G)&=&\phi_{\rm 3}^{-3}
\left[2-e^{-G\phi_{\rm 3}}\{
G^2\phi_{\rm 3}^2(1-\phi_{\rm 3})+2G\phi_{\rm 3}+2\}\right].
\nonumber\\
\label{eq:h}
\ena
$g$ and $h$ change their signs depending 
on $\phi_{\rm 3}$ and $G$. 
Their dependences are shown in Fig.~6. 
If $g(\phi_{\rm 3},G)$ is positive, 
the transition point shifts to higher 
$\chi_{31}$ with an increasing 
$\phi_{\rm 2s}$ and {\it vice versa}. 
Eq.~(\ref{eq:g}) 
for $\phi_{\rm 2s}=0.01$ 
is drawn in the broken curve in Fig.~3 (b). 
Here, we set $\phi_{\rm 3c}=0.2392$ and 
$\chi_{\rm 31t}=0.93$. 
This approximated curve is in aggreement with the numerical solution. 
When $\phi_{\rm 3c}\ll 1$, $g\approx -G(G+2)/2$ and 
it changes its sign 
at $G=0$ and $G=-2$. 
A large negative $h$ enhances the discontinuity of 
the volume change as observed in Fig.~2~(b). 
If $\phi_{\rm 3c}\ll 1$, $h$ is expressed 
as $h\approx G^2(G+3)/3$ and 
it changes its sign at $G=-3$. 
Thus, we can possibly induce the VPT even in 
a gel, which originaly shows a continuous volume change, 
by adding solute of $-G\gg 3$. 
The gap of the VPT disappears 
as $\Delta \phi_{\rm 3t}\propto \sqrt{\phi_{\rm 2st}-\phi_{\rm 2s}}$, 
where $\phi_{\rm 2st}=\{\beta-\beta_{\rm c}^{(0)}\}h^{-1}\phi_{\rm 3c}^{-2}$. 
This disappearing behavior is observed in Fig.~4. 

In this paper, only the numerical 
solutions for $\chi_{12}=0.0$ are presented. 
However, we confirmed that 
the essentially same features are observed 
for any set of $\chi_{12}$ and $\chi_{31}$ 
if the resultant $G$ is the same.

\begin{figure}
\includegraphics[width=84.5mm]{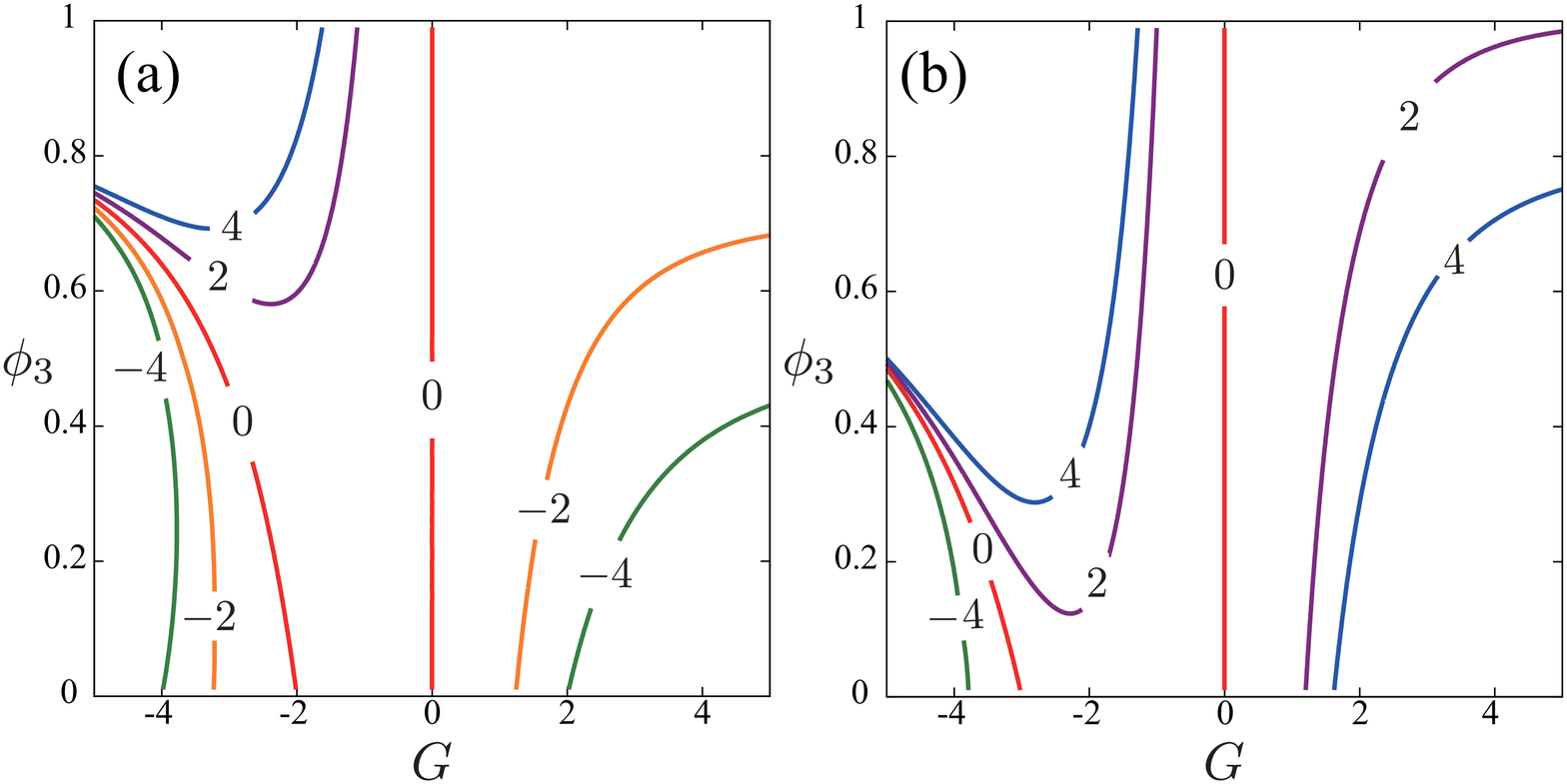}
\caption{
The contours of the functions $g$ [Eq.~(\ref{eq:g})] and 
$h$ [Eq. (\ref{eq:h})] 
are drawn in $G$-$\phi_3$ planes (a) and (b), respectively. 
The transition point is increased linearly with $\phi_{\rm 2s}$ 
in the region of for positive $g$. 
In the region of negative $h$, 
the volume gap of the transition point is increased. 
}
\label{fig6}
\end{figure}

\subsection{The second volume phase transition}

In Fig.~7, we plot 
the swelling curves in a wider range of $\chi_{31}$. 
We set $\phi_{30}=5.0\times 10^{-2}$, $\nu v_0=1.0\times 10^{-2}$, 
$B=0.75$, $\chi_{12}=0.0$ and $\chi_{23}=-16.0$. 
It is indicated that another discontinuous 
volume phase transition can occur 
at $\chi_{31}$ larger than 
$\chi_{\rm 31t}$ corresponding to the first transition. 
We confirmed that 
this second phase transition is observed even if $\nu=0$ 
(or, $\alpha=\beta=0$), 
while the first one disappears. 
This fact indicates that 
this second instability 
is independent of the network elasticity and 
has a physical mechanism different from 
those for well-studied volume phase transitions \cite{Tanaka_PRL_1978}.

Scott reported that 
various types of phase diagrams 
are realized for ternary mixtures 
(polymer solutions in binary mixtures) \cite{Scott_JCP_1949}. 
They can have several critical points, below which 
three phases coexist. 
Analogous to the ternary mixtures, 
we consider that 
the second volume transition observed in Fig.~7 
is attributed to the bulk instability of the mixing free 
energy. 
Since there are some differences between 
ternary mixtures and gels in mixture solvents, 
the Scott's argument cannot be simply applied to 
the gel systems. 
The most important differences are 
that a gel has the elasticity and never exhibits 
a one-phase homogeneous state.

We have not obtained a simple explanation on 
the second transition, since non-linearities, 
which are hard to treat analytically, would play 
imporant roles in the bulk instability. 
In Fig.~7, we also draw the swelling curve 
obtained by Eq.~(\ref{eq:chi31_2}), which 
neglects the non-linearity of $\phi_{\rm 2s}$. 
The approximated curve also 
exhibits the second transition, 
although it does not coincide quantitatively 
with the numerical solutions for large $\phi_3$. 
However, this fact implies that 
the non-linearity of $\phi_{\rm 2}$ has a minor 
contribution to the essential mechanism of the second transition. 

\begin{figure}
\includegraphics[width=42.5mm]{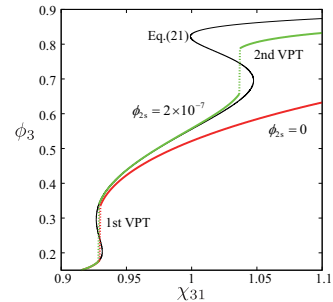}
\caption{
The swelling curves of a gel immersed in a mixture solvent 
in a wider range of $\chi_{31}$. 
We set $\phi_{30}=5.0\times 10^{-2}$, $\nu v_0=1.0\times 10^{-2}$, 
$B=0.75$, and $\chi_{12}=0.0$. 
We dissolve the strongly good solute of $\chi_{23}=-16.0$ and 
$\phi_\mathrm{2s}=2.0\times10^{-7}.$
The black solid lines are calculated by Eq.~(\ref{eq:chi31_2}) .
The second volume phase transition is induced. 
}
\label{fig7}
\end{figure}

As discussed above, 
the gel becomes mechanically unstable 
when $\partial \chi_{31}/\partial \phi_3<0$. 
In Fig.~8, 
we plot $\partial\chi_{31}^{(0)}/\partial \phi_3$ 
and $-h(\phi_3,G)\phi_{\rm 2s}$, 
where $\chi_{31}^{(0)}(\phi_3)$ is the swelling curve 
of the gel in the solvent of the first component. 
In the range satisfying 
$\partial\chi_{31}^{(0)}/\partial \phi_3<-h(\phi_3,G)\phi_{\rm 2s}$, 
the gel would exhibit a discontinuous volume change. 
It is indicated that 
strong non-linearity of $h(\phi_3,G)$ with respect to 
$\phi_3$ is a possible origin of the second transition. 

\begin{figure}
\includegraphics[width=42.5mm]{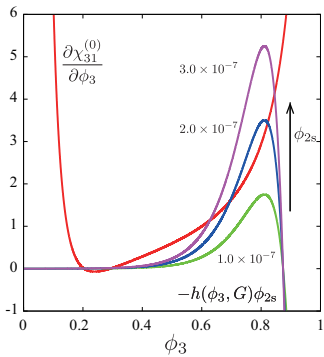}
\caption{
The theoretical curves of 
$\partial \chi_{31}^{(0)}/\partial \phi_3$ and $-h(\phi_3,G)\phi_{\rm 2s}$ 
are shown. 
When $\partial \chi_{31}^{(0)}/\partial \phi_3<-h(\phi_3,G)\phi_{\rm 2s}$, 
the gel becomes mechanically unstable, showing the 
second volume phase transition. 
}
\label{fig8}
\end{figure}

\section{Summary and remarks}

We studied the volume phase transition of gels immersed in binary mixtures, 
based on the three-component Flory-Rehner model. 
Assuming that the volume fraction of the second solvent component is small, 
we reformulated the Flory-Rehner model into a simple model with 
a new parameter $G(=\chi_{23}-\chi_{31}-\chi_{12})$. 
From numerical solutions and the simplified theory, 
we found the following behaviors of the volume phase trasition of the gel. 

(i) 
When the selectivity of the second component to the polymer network is 
small, the composition of the mixture solvent in the gel is close to 
that out of the gel. 
Here, the single liquid approximation works well. 
The renormalized interaction parameter between the solvent and 
the network depends linearly on the composition of the binary mixture. 
Thus, we can swell the gel by dissolving the additive good to the network, 
whereas, the gel is shrunken when the poor additive is dissolved. 

On the other hand, the difference of the composition becomes large 
when the additive selectivity is strong. 
Owing to the non-lineariry of the composition difference, 
the gel tends to be shrunken as the additive concentration 
is increased, regardless of whether it is good or poor. 
From the simplified theory, 
the dependence of the transition point on the addtive concentration 
is found to be proportinal to $\partial \chi_{\rm 31t}/\partial 
\phi_{\rm 2s}\approx g(\phi_{\rm 3c},\bar{G})$, 
which is approximated as $g\approx -\bar{G}(\bar{G}+2)/2$ 
for a hyper swollen gel. 

(ii) 
By dissolving the strongly poor additive, 
we can extinguish the discontinous volume phase transition. 
Furthermore, we can induce the volume phase transition 
in the gel, which does not exhbit the discontinous gap in the 
absence of the additive, by dissolving the strongly 
good additive. 
In the vicinity of the (first) tri-critical point, 
this behavior is well described by the simplified theory 
as shown in Fig.~6 (b).

(iii) 
Far from the transition point, 
we found that another volume phase transition can occur. 
This second instability is observed even if the elasticity 
is negligible. 
The transition is caused by the mixing instability. 
Our theory indicates that the non-linearity of 
$\phi_3$, which is well 
described by Eq.~(\ref{eq:h}), does have a major 
contribution to the transition.

Here, we make some remarks about our results.

1) 
If only van der Waals interaction is taken into account, 
the interaction parameters should be positive, 
{\it i.e.,} $\chi_{ij}\ge0$. 
However, a variety of molecular forces, 
such as ion-dipole interaction, hydrogen bonding and hydrophobic interaction, 
would also influence the volume phase transition of the gel. 
When studying their contributions explicitly, 
we have to consider 
the microscopic degrees of the freedom in a more specific manner 
\cite{FTanaka_PRL_2008,Kojima_Macro_2010,Kojima_SM_2012}. 
In this work, 
we assume that 
the macroscpic $\chi$ parameter can have a large negative value, 
by renormalizing these microscopic degrees into them. 
In a future work, we would like to investigate the connections of 
such microscopic interactions and our phenomenological theory. 
Effects of the selective salts 
\cite{Okamoto_PRE_2010} on polyelectrolyte gels 
are also interesting.

2) 
There are a lot of experiments dealing with NIPA gels and 
PAA gels 
to study the effects of the additive on the volume phase transition. 
Most of additives shrink NIPA gels in agreement with our results. 
Since only a few experiments have reported 
the difference of the additive concentrations between 
in and out of the gels \cite{Ishidao_CPS_1994,Iwatsubo_JMSPB_2001}, 
however, 
we cannot verify the quantitative validity of our model yet. 
Further experimental studies are highly desired. 

\begin{acknowledgments}
We acknowledge valuable discussions with Akira Onuki.
This work was supported by the 
JSPS Core-to-Core Program ``International research network for non-equilibrium dynamics of soft matter" and KAKENHI. 
The computational work was carried out using the facilities at the 
Supercomputer Center, Institute for Solid State Physics, University of Tokyo. 
\end{acknowledgments}

%\bibliographystyle{apsrev}
%\bibliography{UematsuAraki_gel}

\begin{thebibliography}{99}

\bibitem{Flory_JCP_1943}
P. J. Flory and J. Rehner, J. Chem. Phys. {\bf 11}, 512 (1943); 
\textit{ibid} 521 (1943).

\bibitem{Dusek_JPSA_1968} 
K. Du\v{s}ek and D. Patterson, J. Polym. Sci. A-2. {\bf 6}, 1209 (1968). 

\bibitem{Tanaka_PRL_1978}
T. Tanaka, Phys. Rev. Lett. {\bf 40}, 820 (1978). 

\bibitem{Tanaka_PRL_1980}
T. Tanaka, D. Fillmore, S.-T. Sun, I. Nishio, G. Swislow, and A. Shah, 
Phys. Rev. Lett. {\bf 45}, 1636 (1980). 

\bibitem{Tanaka_SA_1981} 
T. Tanaka, Sci. Am. {\bf 244}, 124 (1981). 

\bibitem{Hirokawa_JCP_1984}
Y. Hirokawa and T. Tanaka, J. Chem. Phys. {\bf 81}, 6379 (1984). 

\bibitem{Ohmine_JCP_1982} 
I. Ohmine and T. Tanaka, J. Chem. Phys. {\bf 77}, 5725 (1982). 

\bibitem{Ricka_Macro_1984} 
J. Ri\v{c}ka and T. Tanaka, Macromolecules {\bf 17}, 2916 (1984). 

\bibitem{FernandezNieves_JCP_2001}
A. Fern\'{a}ndez-Nieves, A. Fern\'{a}ndez-Barbero, and F. J. de las Nieves, 
J. Chem. Phys. {\bf 115}, 7644 (2001). 

\bibitem{Amiya_JCP_1987} 
T. Amiya, Y. Hirokawa, Y. Hirose, Y. Li, and T. Tanaka, 
J. Chem. Phys. {\bf 86}, 2375 (1987). 

\bibitem{Hirotsu_JCP_1987} 
S. Hirotsu, Y. Hirokawa, and T. Tanaka, 
J. Chem. Phys. {\bf 87}, 1392 (1987). 

\bibitem{Park_Macro_1993} 
T. G. Park and A. S. Hoffman, Macromolecules {\bf 26}, 5045 (1993). 

\bibitem{Inomata_Lang_1992} 
H. Inomata, S. Goto, K. Otake, and S. Saito, Langmuir {\bf 8}, 687 (1992). 

\bibitem{Annaka_JCP_2000}
M. Annaka, K. Motokawa, S. Sasaki, T. Nakahira, H. Kawasaki, H. Maeda, Y. Amo, 
and Y. Tominaga, J. Chem. Phys. {\bf 113}, 5980 (2000). 

\bibitem{Hirotsu_JPSJ_1987} 
S. Hirotsu, J. Phys. Soc. Jpn. {\bf 56}, 233 (1987). 

\bibitem{Kokufuta_Macro_1993} 
E. Kokufuta, Y.-Q. Zhang, T. Tanaka, and A. Mamada, 
Macromolecules {\bf 26}, 1053 (1993). 

\bibitem{Kawasaki_JPC_1996} 
H. Kawasaki, S. Sasaki, H. Maeda, S. Mihara, M. Tokita, and T. Komai, 
J. Phys. Chem. {\bf 100}, 16272 (1996). 

\bibitem{Sasaki_Macro_1997} 
S. Sasaki, H. Kawasaki, and H. Maeda, Macromolecules {\bf 30}, 1847 (1997). 

\bibitem{Dhara_Lang_1999} 
D. Dhara and P. R. Chatterji, Langmuir {\bf 15}, 930 (1999). 

\bibitem{Koga_JPCB_2001} 
S. Koga, S. Sasaki, and H. Maeda, J. Phys. Chem. B {\bf 105}, 4105 (2001). 

\bibitem{FTanaka_PRL_2008} 
F. Tanaka, T. Koga, and F. M. Winnik, 
Phys. Rev. Lett. {\bf 101}, 028302 (2008). 

\bibitem{Ishidao_CPS_1994} 
T. Ishidao, Y. Hishimoto, Y. Iwai, and Y. Araki, 
Colloid Polym. Sci. {\bf 272}, 1313 (1994). 

\bibitem{Vasilevskaya_PSUSSR_1989} 
V. V. Vasilevskaya, V. A. Ryabina, S. G. Starodubtsev, and A. R. Khokhlov, 
Polym. Sci. U.S.S.R. {\bf 31}, 784 (1989). 

\bibitem{Iwatsubo_Macro_1995} 
T. Iwatsubo, K. Ogasawara, A. Yamasaki, T. Masuoka, and K. Mizoguchi, 
Macromolecules {\bf 28}, 6579 (1995). 

\bibitem{Krigbaum_JPS_1954} 
W. R. Krigbaum and D. K. Carpenter, J. Polym. Sci. {\bf 14}, 241 (1954). 

\bibitem{Bristow_TFS_1959} 
G. M. Bristow, Trans. Faraday Soc. {\bf 55}, 1246 (1959). 

\bibitem{Okeowo_Macro_2006} 
O. Okeowa and J. R. Dorgan, Macromolecules {\bf 39}, 8193 (2006). 

\bibitem{Okamoto_PRE_2010} 
R. Okamoto and A. Onuki, Phys. Rev. E {\bf 82}, 051501 (2010). 

\bibitem{Onuki_COCIS_2011} 
A. Onuki and R. Okamoto, 
Curr. Opin. Colloid Interface Sci. {\bf 16}, 525 (2011). 

\bibitem{Onuki_BCSJ_2011} 
A. Onuki, R. Okamoto, and T. Araki, Bull. Chem. Soc. Jpn. 
{\bf 23}, 284113 (2011). 

\bibitem{Scott_JCP_1949} 
R. L. Scott, J. Chem. Phys. {\bf 17}, 268 (1949). 

\bibitem{Flory_book} 
P. J. Flory, {\it Principle of Polymer Chemistry} (Cornell Univ. Press, 
Ithaca, New York, 1953). 

\bibitem{Kojima_Macro_2010} 
H. Kojima and F. Tanaka, Macromolecules {\bf 43}, 5103 (2010). 

\bibitem{Kojima_SM_2012} 
H. Kojima and F. Tanaka, Soft Matter {\bf 8}, 3010 (2012). 

\bibitem{Iwatsubo_JMSPB_2001} 
T. Iwatsubo and T. Shinbo, J. Macromol. Sci.-Physics. B {\bf 40}, 1017 (2001). 

\end{thebibliography}

\end{document}